\documentclass[fleqn,10pt]{wlscirep}
\usepackage[utf8]{inputenc}
\usepackage[T1]{fontenc}
\title{Mechanical Deformation Induced Continuously Variable Emission for Radiative Cooling}

\author[1]{Xiaojie Liu}
\author[1]{Yanpei Tian}
\author[1]{Fangqi Chen}
\author[2]{Alok Ghanekar}
\author[3,4]{Mauro Antezza}
\author[1,5*]{Yi Zheng}
\affil[1]{Department of Mechanical and Industrial Engineering, Northeastern University, Boston, MA 02115, USA.}
\affil[2]{Artech LLC, Morristown, NJ 07960, USA.}
\affil[3]{Laboratoire Charles Coulomb (L2C), UMR 5221 CNRS–Université de Montpellier, F-34095 Montpellier, France.}
\affil[4]{Institut Universitaire de France, 1 rue Descartes, F-75231 Paris Cedex 05, France.}
\affil[5]{Department of Electrical and Computer Engineering, Northeastern University, Boston, MA 02115, USA.}
\affil[*]{y.zheng@northeastern.edu}

\begin{abstract}
Passive radiative cooling drawing the heat energy of objects to the cold outer space through the atmospheric transparent window (8 $\mu$m \textasciitilde \,13 $\mu$m) is significant for reducing the energy consumption of buildings. Daytime and nighttime radiative cooling have been extensively investigated in the past. However, radiative cooling which can continuously regulate its cooling temperature, like a valve, according to human need is rarely reported. In this study, we present a concept of reconfigurable photonic structure for the adaptive radiative cooling by continuously varying the emission spectra in the atmospheric window region. This is realized by the deformation of the one-dimensional PDMS grating and the nanoparticles embedded PDMS thin film when subjected to mechanical stress/strain. The proposed structure reaches different stagnation temperatures under certain strains. A dynamic exchange between two different strains results in the fluctuation of the photonic structure's temperature around a set temperature.
\end{abstract}
\begin{document}

\flushbottom
\maketitle
%
%
\thispagestyle{empty}


\section*{Introduction}

The growing demand for thermal comfort boosts the increase in the consumption of various energy sources for cooling and heating and exerts enormous stress on electricity systems over the world. This also drives up the carbon dioxide emissions and contributes to the problem of global warming. Nearly 20\% of the total electricity was used by air conditioners or electric fans to regulate the temperature of buildings to be comfortable \cite{international2018future}. However, the peak wavelength (\textasciitilde \,9.7 $\mu$m) of blackbody radiation for objects on Earth (\textasciitilde \,300K) coincides with the atmospheric highly transparent window (8 \textasciitilde \,13 $\mu$m) that scarcely absorb infrared thermal radiation. Therefore, terrestrial objects can naturally radiate thermal energy to the outer space (\textasciitilde \,3K) through the atmospheric window and hence lower their temperature, which is called passive radiative cooling \cite{ono2018self}. 

Effective nighttime radiative cooling has been extensively studied for organic and inorganic materials with high infrared emissivity within the atmospheric window \cite{catalanotti1975radiative,granqvist1980surfaces,orel1993radiative}. However, the daytime radiative cooling is highly demanded and a challenge since the solar radiation (ASTM G-173, \textasciitilde\,1000W/m$^2$) is much higher than the potential radiative cooling (\textasciitilde \,100W/m$^2$). If objects absorb only a few percents of solar irradiance, it will counteract the cooling power and heats the objects ultimately. To achieve daytime radiative cooling, a spectrally selective surface which effectively reflects solar irradiance (0.3 $\mu$m \textasciitilde \,2.5 $\mu$m) and strongly emits heat within the infrared region (8 $\mu$m \textasciitilde \,13 $\mu$m) simultaneously is a promising device. Consequently, several metamaterials successfully achieving daytime radiative cooling with an equilibrium temperature below the ambient have been experimentally investigated, such as silica-polymer hybrid metamaterial \cite{zhai2017scalable}, hierarchically porous paint-like materials \cite{mandal2018hierarchically}, and wood-based structural materials \cite{li2019radiative}. Other materials like nanophotonic structures \cite{rephaeli2013ultrabroadband,raman2014passive}, infrared transparent aerogel \cite{leroy2019high}, and polymer nanofiber \cite{wang2020scalable} also provide various alternatives for daytime radiative cooling. These materials pave the way for applications of radiative cooling to energy-saving buildings, energy harvesting, and temperature regulation without energy consumption and achieving sustainable cooling throughout the day.

Although static radiative cooling systems can effectively save energy in summer, the cooling functionality will increase the energy consumption for heating in winter. To overcome this difficulty, a conceptive design of self-adaptive radiative cooling was developed based on phase change material vanadium dioxide (VO$_2$) that can adaptively turn "ON" and "OFF" radiative cooling corresponding to the ambient temperature \cite{ono2018self,wu2017thermal}. Moreover, the phase change temperature of VO$_2$ co-doping with W and Sr can be adjusted around the room temperature by changing W contents \cite{dietrich2017optimizing}. Although these temperature-induced systems can automatically adjust the radiative cooling with ambient temperature, their performances, and applications highly based on the specific phase change temperature under the specific W content. Considering the rigorous manufacturing process and the fixed sole phase change temperature under specific usage scenarios, there are still some limitations on large-scale fabrications and complex practical applications.

\begin{figure*}[!t]
\centering
\includegraphics[width=1\textwidth]{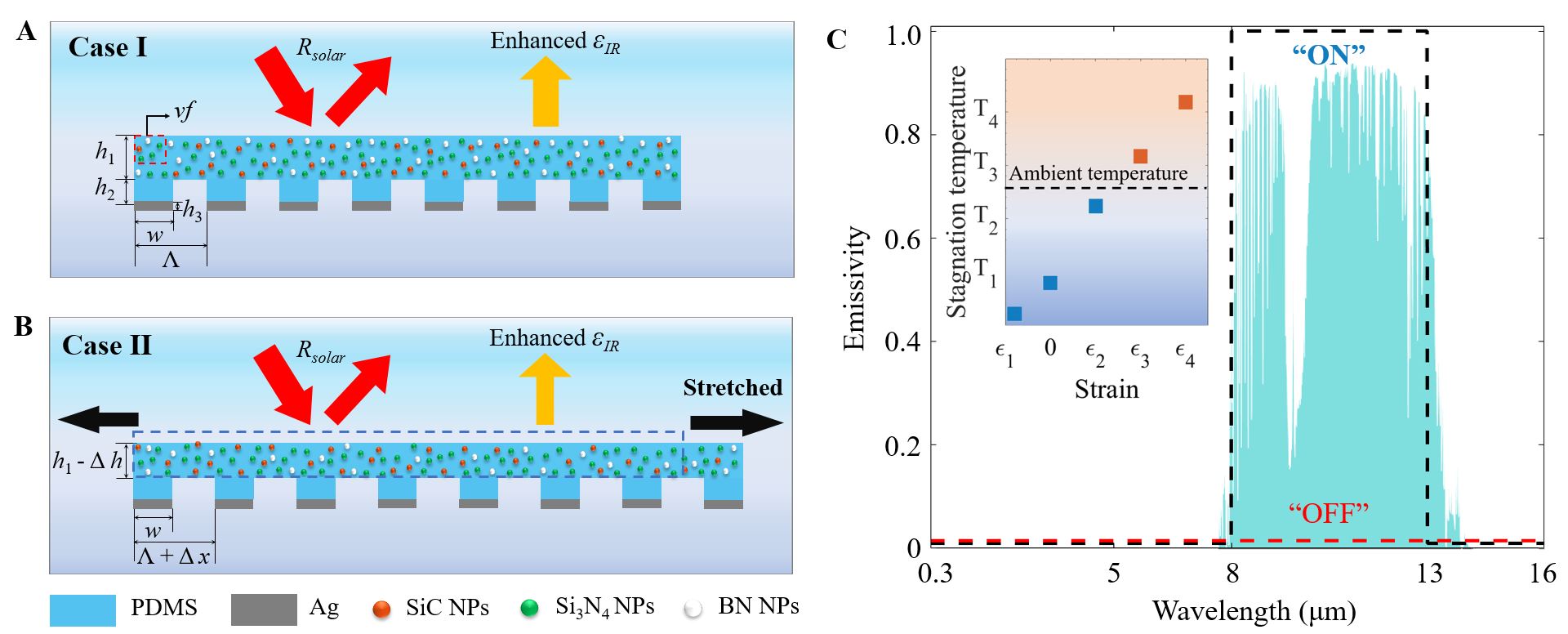}
\caption{ \label{fig:schematic_cooling} (\textbf{A})  The reconfigurable metamaterials consists of a PDMS layer (thickness, \textit{h}$_1$) embedded with three nanoparticles: SiC, Si$_3$N$_4$, and BN (volume fraction, \textit{vf}$_{SiC/Si_3N_4/BN}$). 1-D rectangular grating (period ($\Lambda$), width (\textit{w}), filling ration ($\phi$)) of PDMS (\textit{h}$_2$) coated with silver (Ag, \textit{h}$_3$) thin film is at the back of the top PDMS layer. Case I: the structure has high \textit{R}$_{solar}$ that reflects most of the solar irradiance, and has high $\epsilon_{IR}$ in atmospheric window region that can radiate heat out to the universe when it is released. (\textbf{B}) Case II: it keeps unchanged high \textit{R}$_{solar}$, while the $\epsilon_{IR}$ is reduced since the PDMS layer gets elongated and thinner and the period of the 1-D PDMS layer is increased when the grating is stretched, but the width of the Ag layer does not change. (\textbf{C}) Schematic showing the concept of mechanical deformation induced radiative cooling.
}
\end{figure*}

Here, we conceptually propose a system of a reconfigurable nanophotonic structure for mechanical deformation induced radiative cooling basing on the continuously variable emission in the atmospheric window to attain diverse desired stagnation temperatures by continuous deformation adjustment according to ambient temperature.

\section*{Results}

The reconfigurable structure consists of a PDMS layer embedded with multispecies of nanoparticles on top of the 1-D PDMS grating coated by a silver thin film (Fig. \ref{fig:schematic_cooling}\textbf{A}). The emissivity spectra in the atmospheric window of this structure are continuously tunable by the mechanical deformation of the top PDMS thin film and PDMS grating periods to stabilize at a certain temperature when subjected to a mechanical strain (Fig. \ref{fig:schematic_cooling}\textbf{B}). We theoretically prove that the emissivity properties of the proposed system under different strains are angular-independent which is important in real applications. Theoretical analysis also shows that this system can maintain itself at a set temperature by mechanical deformation which could be potentially applied to thermal regulations for different applications, such as outdoor vehicles, buildings, and greenhouses.

Figure \ref{fig:schematic_cooling} \textbf{C} introduces the concept of the mechanical deformation induced radiative cooling. The basic principle of continuous temperature adjustment is that the thickness of the top nanoparticles embedded PDMS layer and the period of the Ag coated PDMS gratings are changed with mechanical deformation, which induces the corresponding change in emissivity of the structure with the atmospheric window (8 \textasciitilde \,13 $\mu$m). This structure, like a valve, can continuously regulate its opening when subjected to different strains. The emissivity in the atmospheric window is a function of the strain. The higher the strain, the lower the emissivity. Furthermore, different strains correspond to different stagnation temperatures, that is a small strain yields a stagnation temperature below the ambient temperature, while a large strain represents a stagnation temperature above the ambient (the inset of the Fig. \ref{fig:schematic_cooling} \textbf{C}). 

\begin{figure*}[!t]
\centering
\includegraphics[width=1\textwidth]{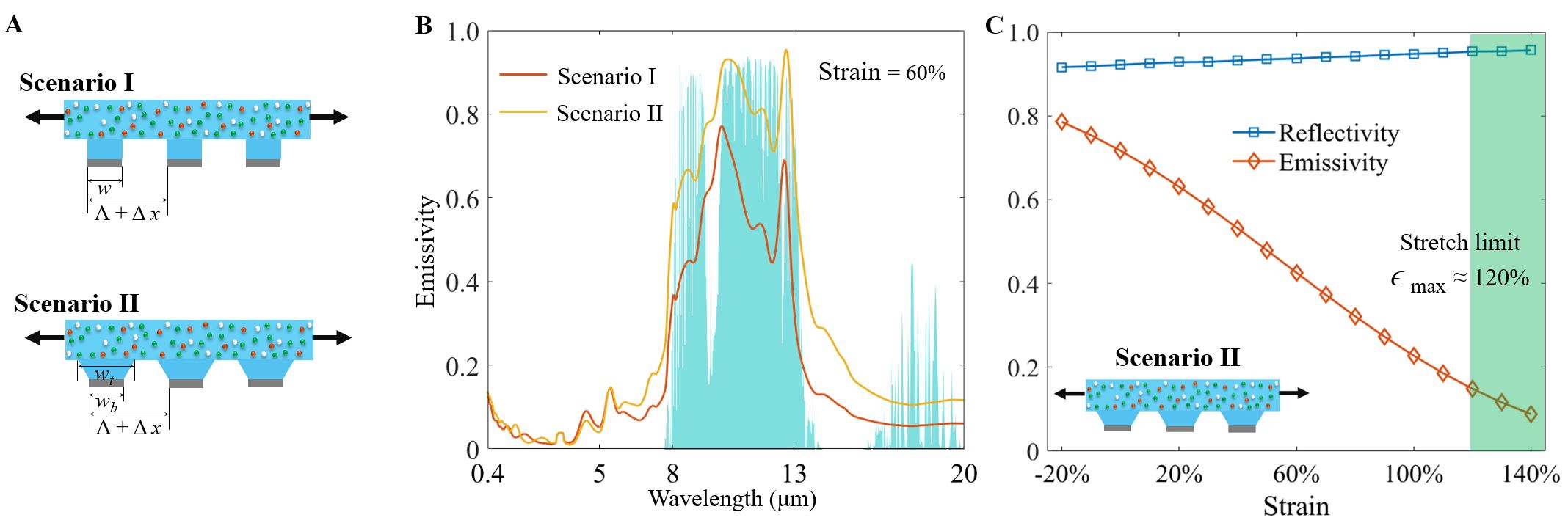}
\caption{ \label{fig:diff_scenario} (\textbf{A}) Two scenarios of the reconfigurable nanophotonic structure under stretching. Top: scenario I (constant \textit{w}) for the ideal stretching or compression due to the mechanical strain. Bottom: scenario II (constant \textit{w}$_t$) for the real stretching situation. (\textbf{B}) The spectral emissivity of the reconfigurable metamaterials for different scenarios under 60\% strain. (\textbf{C}) The strain dependent overall reflectively and emissivity for scenario II. The overall \textit{R}$_{solar}$ is calculated from 0.4 $\mu$m to 2.5 $\mu$m, and the $\epsilon_{IR}$ is calculated from 8 $\mu$m to 13 $\mu$m.
}
\end{figure*}

To realize such functionality, we employ an elastomer, PDMS, as the valve's component to forms reconfigurable metamaterials and proposed a nanophotonic structure with a PDMS layer embedded with three nanoparticles: SiC, Si$_3$N$_4$, and BN. 1-D PDMS grating layer coated with Ag thin film adheres to the top PDMS layer. When subjected to mechanical deformation of $\Delta$\textit{x}, The PDMS stretches and the grating period ($\Lambda$) and filling ratio ($\phi$) increase. We assume that the Ag grating strips (width, \textit{w}) do not undergo any deformation as it has a much higher Young's modulus (69) than the PDMS (0.5). Therefore, the new grating period of the stretched structure is $\Lambda$ + $\Delta$\textit{x} and the new filling ratio is \textit{w}/($\Lambda$ + $\Delta$\textit{x}). The thickness of the top PDMS layer, \textit{h}$_1$, also decrease to be (\textit{h}$_1$ - $\Delta$\textit{h}), as shown in Fig. \ref{fig:schematic_cooling} \textbf{B}. The PDMS strongly absorbs infrared light when its thickness is above 1 $\mu$m since its extinction coefficient ($\kappa$) has absorption peaks from 7 $\mu$m to 13 $\mu$m \cite{querry1987optical}. If we increase \textit{h}$_1$ above 10 $\mu$m, it emissivity will increase to 0.9 but its spectral selectivity loses, so we keep its thickness around 1 $\mu$m and introduce three nanoparticles (SiC, Si$_3$N$_4$, and BN) to increase the emissivity only within the atmospheric window. These three nanoparticles have separate extinction coefficient peaks from 7 $\mu$m to 13 $\mu$m (SiC: 12.8 $\mu$m \cite{larruquert2011self}; Si$_3$N$_4$: 8.5 $\mu$m and 12.5 $\mu$m \cite{liu2019spectral}; BN: 7.09 $\mu$m and 12.45 $\mu$m \cite{liu2019spectral}). This increases the emissivity in the atmospheric window but does not for the rest wavelength range. The PDMS grating strips serve as a transition layer between the top PDMS layer and the bottom Ag layer. Although the strain of the structure increase to the PDMS limits (120\%), the Ag layer can still keep undeformed. Since Ag is highly reflective from 0.37 $\mu$m to 20 $\mu$m \cite{ciesielski2017controlling}, so the Ag grating layer can be regarded as opaque to both the infrared and visible light considering its thickness we used, \textit{h}$_3$= 400 nm and the small period, $\Lambda$= 40 nm compared to the wavelength range consider here (0.37 $\mu$m to 20 $\mu$m), that is, the Ag grating layer serves as a thin film to reflect all the incident light, even under 120\% strain. Since the PDMS layer is transparent in the solar wavelength region as it has a negligible extinction coefficient \cite{zhang2020complex}, the proposed structure is highly reflective in the solar region.

\begin{figure*}[!ht]
\centering
\includegraphics[width=0.9\textwidth]{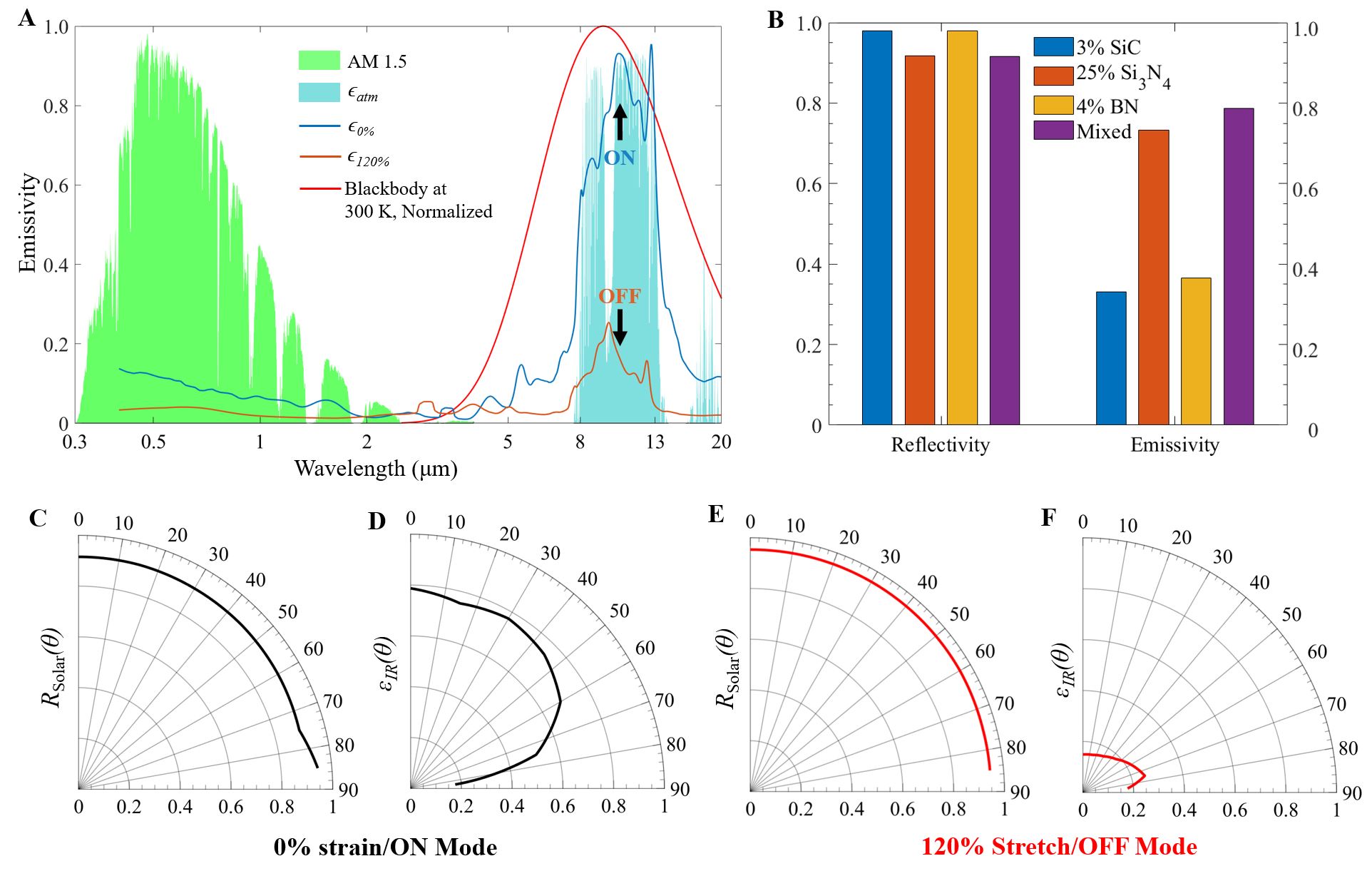}
\caption{ \label{fig:mainfigure} (\textbf{A}) Spectral emissivity ($\epsilon$ = 1 - \textit{R}) of the reconfigurable nanophotonic structure in the original state and under a strain of 120\% displayed with the normalized ASTM G173 solar spectrum (AM 1.5), the infrared atmospheric transparent window and the normalized blackbody spectrum at 300K. (\textbf{B}) The overall \textit{R}$_{solar}$ and $\epsilon_{IR}$ of the proposed structure embedded with only the single species of SiC, $Si_3N_4$, and BN and the three mixed nanoparticles. The nanophotonic structure's \textit{R}$_{solar}$ $(\theta)$ (\textbf{C}) and $\epsilon_{IR}$ ($\theta$) (\textbf{D}) across various angle of incident (AOI) result in high hemispherical \textit{R}$_{solar}$ and $\epsilon_{IR}$ at original state ("ON" mode), and \textit{R}$_{solar}$ $(\theta)$ (\textbf{E}) under a strain of 120\%. The low $\epsilon_{IR}$ ($\theta$) (\textbf{F}) across angles shows the "OFF" mode with a low hemispherical $\epsilon_{IR}$ under 120\% strain. 
}
\end{figure*}

The hemispherical emissivity of the reconfigurable nanophotonic structure can be expressed as \cite{ghanekar2015role}:

\begin{equation}
\label{eq:1}
\epsilon(\omega)=\frac{c^{2}}{\omega^{2}}\int_0^{\omega/c}dk_{\rho}k_{\rho}\sum_{\mu=s,p}(1-|\widetilde{R}_{h}^{(\mu)}|^{2}-|\widetilde{T}_{h}^{(\mu)}|^{2}) 
\end{equation}

where c is the speed of light in vacuum, $\omega$ is the angular frequency and $k_{\rho}$ is the magnitude of inplane wave vector. $\widetilde{R}_{h}^{(\mu)}$ and $\widetilde{T}_{h}^{(\mu)}$ are the polarization dependent effective reflection and transmission coefficients which can be calculated using the recursive relations of Fresnel coefficients of each interface \cite{chew1995waves}. The dielectric functions can be related to real $(n)$ and imaginary $(\kappa)$ parts of refractive index as $\sqrt{\varepsilon}=n+j\kappa$. Dielectric functions of the materials (PDMS, SiC, Si$_3$N$_4$, BN and Ag) utilized in this work are taken from literature \cite{zhang2020complex,querry1987optical,palik1998handbook,luke2015broadband,neuner2010midinfrared,yang2015optical}. The Bruggeman effective medium theory is employed to predict the dielectric function of the nanoparticles embedded PDMS thin film composite. Here, the diameter of these three nanopartiles are confined to be 80 nm that is much smaller than the shortest wavelength of interest (400 nm) and the thickness of the PDMS layer.  Besides, the sum volume fractions of these nanoparticles are below 33\% (the maximum volume fraction limit of Bruggeman effective medium approximation) \cite{choy2015effective}. Therefore, the effective dielectric function of multispecies of nanoparticles can be calculated:
\begin{equation}
\label{eq:2}
\sum_{i} \eta_{i}\left(\frac{\epsilon_{i}-\epsilon}{\epsilon_{i}+2 \epsilon}\right)=0
\end{equation}
where $\eta_{i}$ is the volume fraction of different nanoparticles, $\epsilon_{i}$ stands for the dielectric function of different nanoparticles. $\epsilon$ is the dielectric function of the matrix. As our design involves a 1-D grating structure of PDMS, second order approximation of effective medium theory was used to obtain the effective dielectric properties given by \cite{raguin1993antireflection}: 
\begin{subequations}
\begin{equation}
\label{eq:3_1}
\varepsilon_{TE,2}\!=\!\varepsilon_{TE,0}\!\left[\! 1\!+\!\frac{\pi^2}{3}\!\left(\frac{\Lambda}{\lambda}\right)^2\!\phi^2(1-\phi)^2\frac{(\!\varepsilon_{A}\!-\!\varepsilon_{B}\!)^2}{\varepsilon_{TE,0}}\right]
\end{equation}

\begin{equation}
\label{eq:3_2}
\!\varepsilon_{TM,2}\!=\!\varepsilon_{TM,0}\!\left[\!1\!+\!\frac{\!\pi^2}{\!3}\!\left(\!\frac{\!\Lambda}{\!\lambda}\!
\right)^2\!\phi^2(\!1\!-\!\phi)^2 (\!\varepsilon_{\!A}\!-\!\varepsilon_{\!B}\!)^2\varepsilon_{TE,0}\!\left(\!\frac{\varepsilon_{TM,0}}{\varepsilon_{\!A}\varepsilon_{\!B}}\!\right)^2\right]
\end{equation}
\end{subequations}
where $\varepsilon_{A}$ and $\varepsilon_{B}$ are dielectric functions of two media (PDMS and vacuum) in surface gratings. The expressions for zeroth order effective dielectric functions $\varepsilon_{TE,0}$ and $\varepsilon_{TM,0}$ are given by \cite{glytsis1992high,raguin1993antireflection}:
We choose grating period $\Lambda$= 40 nm that is much smaller than the shortest wavelength (400 nm). 

In order to better fit the practical application scenarios, here, two possible scenarios of the deformation for the transition PDMS grating layer are considered. In fig. \ref{fig:diff_scenario} \textbf{A}, scenario I (top photonic structure) shows the ideal scenario. The \textit{w} of PDMS grating remains unchanged when the period elongates from $\Lambda$ to $\Lambda$ + $\Delta$\textit{x}. However, the PDMS grating layer must undergo deformation to some extent. Hence, we assume that the bottom width, \textit{w}$_b$ keeps unchanged, while the top width, \textit{w}$_t$ elongates with the same strain to the top PDMS layer. Therefore, the PDMS grating strips become an isosceles trapezoid subjected to the mechanical stain, which represents the practical situations, like scenario II. To illustrate the difference between scenarios I and II, the emissivity spectra of the structure under 60\% strain are calculated and shown in Fig. \ref{fig:diff_scenario} \textbf{B}. The difference between the two scenarios cannot be negligible in the infrared region, so scenario II is adopted for the following analysis. The spectral emissivity of scenario I is higher than scenario II's, and the reason is that the PDMS grating strip in scenario II fills more in the vacuum space than the scenario I when subjected to the same mechanical strain. This increases the infrared absorptance over the atmospheric window region. For the deformation of the PDMS grating layer, we divide the 1-D strips into multiple layers of rectangular gratings with decreasing filling ration from top to bottom. Here, we take 100 layers in calculations which is enough to get converged. Fig. \ref{fig:diff_scenario} \textbf{C} shows the strain-dependent reflectivity in the wavelength range of 0.4 $\mu$m to 2.5 $\mu$m and emissivity in the wavelength range of 8 $\mu$m to 13 $\mu$m for scenario II. The \textit{R}$_{solar}$ increases slightly with strain and the $\epsilon_{IR}$ drops abruptly as the strain increases, for example, the $\epsilon_{IR}$ at 100\% (0.23) strain is only equivalent to 32\% of the $\epsilon_{IR}$ at the original state. This is because the thickness of the top PDMS layer decrease and the incident infrared light travels less in the top PDMS thin film. We confine the strain less than 120\% since the PDMS film will fracture around that strain \cite{bohmer2018constructing}.

After the optimization of variables \textit{h}$_{1}$, \textit{h}$_{2}$, \textit{h}$_{3}$, $\Lambda$, $\Phi$ and \textit{vf} of SiC, Si$_3$N$_4$, and BN, we get the optimal configuration with \textit{h}$_{1}$= 1100 nm, \textit{h}$_{2}$= 100 nm, \textit{h}$_{3}$= 400 nm, $\Lambda$= 40 nm, $\Phi$= 0.6, \textit{vf}$_{SiC}$= 3\%, \textit{vf}$_{Si_3N_4}$= 25\%, and \textit{vf}$_{SiC}$= 4\%. Figure \ref{fig:mainfigure} \textbf{A} shows the spectral emissivity of the reconfigurable nanophotonic structure at two limit states: original state (0\% strain) and 120\% strain. Both the original and stretched structure show high reflectivity in the solar irradiance region, while the original one has relatively high emissivity over atmospheric window that represents the complete "ON" of radiative cooling valve. The stretched structure has an overall 0.15 emissivity from 8 $\mu$m to 13 $\mu$m and stands for the entire "OFF" features of the valve. Besides, both the "ON" and "OFF" states have low absorptivity in the rest wavelength region (5 $\mu$m \textasciitilde \,8 $\mu$m and 13 $\mu$m \textasciitilde \,20 $\mu$m) for thermal radiation which avoid the absorption of heat from the ambient environment. The structure of three nanoparticle inclusions has similar \textit{R}$_{solar}$ while has relative higher $\epsilon_{IR}$ over the atmospheric window compared with the structure of the single nanoparticle inclusions (Fig. \ref{fig:schematic_cooling} \textbf{B}). The high \textit{R}$_{solar}$ ($\theta$) ensures excellent reflection of sunlight from all angles of incidences (Figs. \ref{fig:mainfigure} \textbf{C} and \ref{fig:mainfigure} \textbf{E}, angle-averaged emissivity: 0.92 and 0.95), and the high $\epsilon_{IR}$ ($\theta$) (Fig. \ref{fig:mainfigure} \textbf{D}, angle-averaged emissivity: 0.6324) of the complete "ON" state from 0$^\circ$ to 60$^\circ$ leads to a hemispherical high $\epsilon_{IR}$ resulting in a good radiative cooling feature. However, the low $\epsilon_{IR}$ ($\theta$) of the entire "OFF" state from 0$^\circ$ to 85$^\circ$ (Fig. \ref{fig:mainfigure} \textbf{F}, angle-averaged emissivity: 0.18) yields the low radiative cooling ability. Moreover, the states between the entire "ON" and "OFF" state represent different emissivity in the atmospheric window corresponding to different strains. 

\begin{figure*}[!t]
\centering
\includegraphics[width=1\textwidth]{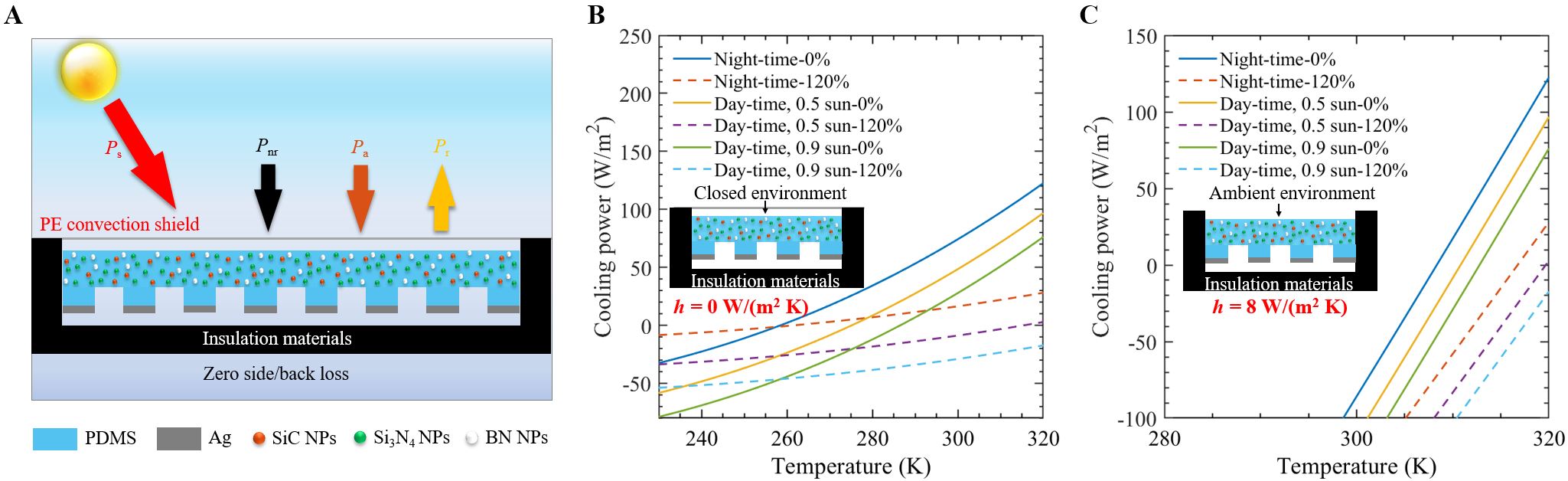}
\caption{ \label{fig:coolingpower} (\textbf{A}) Schematic drawing of the thermal characterization setup used in the thermal performance analysis. (\textbf{B}) Calculated net cooling power of the reconfigurable nanophotonic structure at a different strain as a function of its temperature at night-time and day-time under 0.5 sun (0.5 * AM 1.5 illumination) and 0.9 sun (0.9* AM 1.5 illumination) with (\textbf{B}) or without (\textbf{C}) polyethylene (PE) convection shield.
}
\end{figure*}

\section*{Discussions}
The thermal performance analysis of the reconfigurable metamaterials is evaluated by solving the energy balance equation (Figure \ref{fig:coolingpower} \textbf{A}):

\begin{equation}
\label{eq:4}
P_{\text {net}}=P_{\text {r}}\left(T_{\text {}}\right)-P_{\text {nr}}\left(T_{\text {a}}, T_{\text {}}\right)-P_{\text {a}}\left(T_{\text {a}}\right) -P_{\text {s}}\left(T_{\text {}}\right)
\end{equation}

We supposed that the backside of the self-adaptive photonic structure is insulated, and only the energy transfer between the top surface of the structure, ambient, and outer space is considered. Here, $P_r$ is the radiative cooling power of the structure, $P_{nr}$ is the non-radiative power from the ambient, $P_a$ is the incident thermal radiation power from the ambient, $P_s$ stands for the incident solar power absorbed by the structure, $T_a$ means the temperature of ambient air, and $T$ presents the temperature of the structure.
$P_r$ can be determined as follows:

\begin{equation}
\label{eq:5}
P_{\text {r}}\left(T_{\text {}}\right)=\int_{\text {0}}^{\infty} \mathrm{d} \lambda I_{B B}\left(T_{\text {}}, \lambda\right) \varepsilon\left(\lambda, \theta, \phi, T_{\text {}}\right)
\end{equation}
where, $I_{BB} (T, \lambda)$ = $2hc^{2}{\lambda^{-5}}$ $\exp( hc/\lambda k_{B}T-1)^{-1}$ defines the spectral radiance of blackbody at a certain temperature. where \textit{h} is the Planck's constant, $k_{B}$ is the Boltzmann constant, and $\lambda$ is the wavelength. $\epsilon(\lambda, \theta, \phi, T_{cooler})$ = $\frac{1}{\pi}$$\int_0^{2\pi}{\rm d}\phi \int_0^{\pi/2} \epsilon_{\lambda}\cos\theta \sin \theta {\rm d}\theta $ is the temperature-dependent emissivity of the structure \cite{zhang2007nano}. Here, the emissivity measured at room temperature (298 K) is taken into simulation, since it is assumed that the temperature variations of the structure affect little on emissivity. $\theta$ and $\phi$ are the azimuthal and latitudinal angles, respectively.

The non-radiative heat transfer between the structure and ambient air is given by:
\begin{equation}
\label{eq:6}
P_{nr}(T_{a}, T) = h(T_{a}-T_{})
\end{equation}
$h$ is the nonradiative heat transfer coefficient ranging from 2 to 8 Wm$^{-2}$K$^{-1}$ \cite{ono2018self}. Here $h$ = 8 Wm$^{-2}$K$^{-1}$ is set as natural air convection heat transfer to the structure. The absorbed power of the incident thermal radiation from atmosphere $P_{a} (T_{a})$ is given by:
\begin{equation}
\label{eq:7}
\begin{aligned}
P_{a}(T_{a}) = \int_0^\infty & {\rm d}\lambda I_{BB} (T_{a},\lambda)\epsilon (\lambda, \theta, \phi, T_{}) \epsilon (\lambda, \theta, \phi)
\end{aligned}
\end{equation}
The absorptivity of the atmosphere, $\epsilon (\lambda, \theta, \phi)$, is given by 1-$\tau(\lambda, \theta, \phi)$. Here $\tau(\lambda, \theta, \phi)$ is the transmittance value of atmosphere obtained from MODTRAN4 \cite{berk1999modtran4}. Solar irradiation absorbed by the radiative cooler $P_{s}(T)$ is given by:
\begin{equation}
\label{eq:8}
P_{s}(T) = \int_0^\infty {\rm d}\lambda I_{AM 1.5} (\lambda) \epsilon (\lambda, \theta_{s}, T)
\end{equation}
Here, $I_{\mathrm{AM} 1.5}(\lambda)$ is the spectral irradiance intensity of solar irradiation at AM 1.5. $\epsilon (\lambda, \theta_{sun}, T_{cooler})$ is the temperature-dependent emissivity of radiative cooler. The integration is taken from 0.3 $\mu$m to 2.5 $\mu$m, which cover 97\% of the solar incident power. The time-dependent temperature variations of the structure can be obtained by solving the following equation:
\begin{equation}
\label{eq:9}
C \frac{dT}{dt}= P_{net}(T, T_{a})
\end{equation}
Since Ag has relatively high thermal conductivity (406 W/m K) and the Ag grating strips' thickness is only 400 nm, so the thermal resistance is negligible. The heat capacitance of the reconfigurable photonic structure, $C$, consists of the PDMS thin film and PDMS grating strip with a thickness of 1200 nm ($h_1$ + $h_2$).

\begin{figure*}[!t]
\centering
\includegraphics[width=0.8\textwidth]{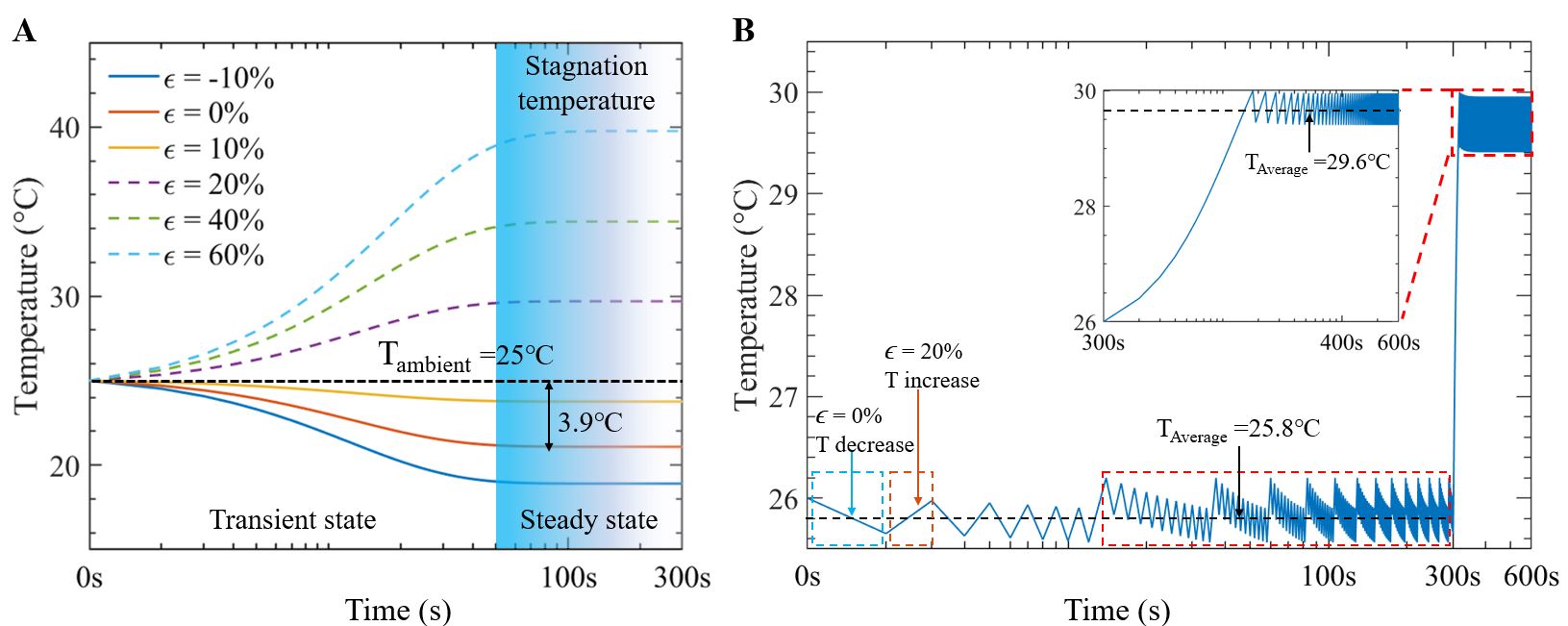}
\caption{ \label{fig:Self_adaptive} (\textbf{A}) Stagnation temperature of the reconfigurable metamaterials as a function of strains showing the structure has cooling or heating abilities under different strains. (\textbf{B}) Transient temperature variations of the structure when subjected to dynamic mechanical strain showing it can control its temperature around a critical temperature.
}
\end{figure*}
We present the net cooling power as a function of the structure's temperature without (Fig. \ref{fig:coolingpower} \textbf{B}) and with (Fig. \ref{fig:coolingpower} \textbf{C}) the influence of non-radiative heat transfer, respectively. Figs. \ref{fig:coolingpower} \textbf{B} and \ref{fig:coolingpower} \textbf{C} show that the structure has a larger net cooling power in a closed environment (h = 0 W m$^{-2}$ K$^{-1}$) than the one that is open to the ambient environment (h = 8 W m$^{-2}$ K$^{-1}$) at any temperature for different strains. The net night-time radiative cooling is higher than the day-time's since the absorbed solar irradiance neutralizes part of the cooling power that the structure radiates out to the outer space in the daytime. The net cooling power of the original structure is higher than the stretched one's (120\%) whether it is open to ambient or not, and most of the stretched one's cooling power is negative, that is, it increases the temperature of the structure. Therefore, the original and stretched state of the reconfigurable structure can be regarded as the complete "ON" and "OFF" states. The temperature of the closed environment in which the net cooling power is positive is lower than the open environment (night-time: 260K-0\% strain, 307K-0\%; day-time (0.9 suns): 287K-0\%, 313K-0\%), this is also for the stretched state because the PE convection shield eliminates the non-radiative heating power from the ambient. Therefore, the closed environment is better for a lower desired temperature, while the open environment case is suitable for a higher expected temperature. 

The stagnation temperature responses of the continuously adaptive cooling structure under various strains are presented in Fig. \ref{fig:Self_adaptive} \textbf{A} by solving Eq. \ref{eq:9} using spectra under different strains. For each strain, both the structure and the ambient is assumed to be 25$^\circ$C and we set $h$ = 0 W m$^{-2}$ K$^{-1}$ and $I_{solar}$ = 1 sun. When the strain is below 20\%, the net cooling power of the reconfigurable structure is positive, its temperature decrease as time evolves, and eventually reaches a stagnation temperature which below the ambient after the 50s (-10\% strain $\rightarrow$ 6.1$^\circ$C temperature decrease, 0\% strain $\rightarrow$ 3.9$^\circ$C temperature decrease, 10\% strain $\rightarrow$ 1.24$^\circ$C temperature decrease). While the strain is above 20\%, the system is approaching to the complete "OFF" state, the negative radiative cooling heats the structure up and reaches an equilibrium temperature that is above the initial temperature (20\% strain $\rightarrow$ 4.69$^\circ$C temperature increase, 40\% strain $\rightarrow$ 9.4$^\circ$C temperature increase, 60\% strain $\rightarrow$ 14.76$^\circ$C temperature increase). 

Finally, we simulate the transient temperature variations of the structure as a function of time when subjected to dynamic mechanical strains to keep at a set temperature(Fig. \ref{fig:Self_adaptive} \textbf{B}). The system is under an environment with $h$ = 0 W m$^{-2}$ K$^{-1}$, $I_{solar}$ = 1 sun and 26 $^\circ$C of the ambient temperature. The initial temperature of the structure is assumed to be 26 $^\circ$C which gives humans thermal comfort. We use the spectra of the structure at the original state and 20\% strain into the calculation. The set temperature of this structure in the first 300s is 26 $^\circ$C. When the structure is in the original state, the radiative cooling feature is total on, and then the structure's temperature drops below 26 $^\circ$C from 0s to 1s. The temperature of the structure goes up from 1s to 2s, since the structure is stretched by 20\%. The transient temperature of the structure keeps dynamically changing between 25.7$^\circ$C and 26.05$^\circ$C after the 15s with an average temperature of 25.8 $^\circ$C. This shows that this system can control its temperature in a narrowband around the set temperature. When we change the set temperature at 300s to be 30$^\circ$C, the structure's temperature can go up to 30$^\circ$C and then stay around that point. The average temperature from 315s to 600s is 29.6$^\circ$C. This shows our proposed structure has a quick adjustment ability.

Above all, we have presented a conceptive reconfigurable nanophotonic design of mechanical-induced radiative cooling that can continuously adjust the radiative cooling when subjected to different mechanical strains. A PDMS thin film and grating strips allow the reversible stretching of the structure. Deformation of the PDMS thin film and PDMS gratings leads to a change of the thin film thickness and filling ratio of PDMS grating strips, and hence, the spectral emissivity of the structure over the atmospheric window can be actively and continuously changed. Compared with other self-adaptive radiative cooling for fixed critical temperature, our designs have various stagnation temperature under different strains, which give us more options for different engineering applications. Moreover, strains above 20\% can turn the radiative cooling to heating and fluctuational temperature control can be achieved with the dynamic exchange between 0\% and 20\% strain. This work verifies that the elastic materials have the potential to be applied in the mechanical deformation induced radiative cooling. As a proof of concept demonstration, we use elastic polymer, PDMS, as the matrix materials, and other transparent soft polymers, like silica gel, can also serve as the alternative matrix. Such designs can be potentially applied in a series of applications, such as energy-saving buildings, textiles, and automobiles for energy-saving and thermal comfort enhancing.





\section*{Acknowledgements}

This project is supported by the National Science Foundation through grant number
CBET-1941743.

\section*{Author contributions statement}
X.L. and Y.Z. develop the model. X.L. and Y.T. do the calculation and write the manuscript with help from all other authors. A.G., F.C. and M.A. contribute to the development of ideas and approaches. All authors provide critical feedback and help revise the final version of the manuscript. Y.Z. supervises this project.

\section*{Additional information}
The authors declare no conflict of interest.



\bibliography{Xiaojie}


\end{document}